\documentstyle[aasms4,psfig]{article}

\slugcomment{Accepted for publication in Astrophysical Journal Letters}

\lefthead{Greaves et al.}
\righthead{A dust ring around Epsilon Eridani}

\begin{document}

\title{A dust ring around Epsilon Eridani: \\
analogue to the young Solar System}

\author{J.S. Greaves$^*$, W.S. Holland, G. Moriarty-Schieven and 
T. Jenness}
\affil{Joint Astronomy Centre, 660 N. A`oh\={o}k\={u} Place, 
University Park, Hilo, HI~96720 \\
{\em $^*$offprint requests: jsg@jach.hawaii.edu}}
\and
\author{W.R.F. Dent}
\affil{Royal Observatory, Blackford Hill, Edinburgh EH9 3HJ, U.K.}
\and
\author{B. Zuckerman, C. McCarthy and R.A. Webb}
\affil{Department of Astronomy, University of California, Los Angeles, Los
Angeles, CA 90095}
\and
\author{H.M. Butner}
\affil{SMTO, University of Arizona, Tucson, AZ 85721}
\and
\author{W.K. Gear}
\affil{Mullard Space Science Laboratory, University College London, Holmbury
St. Mary, Dorking, Surrey RH5 6NT, U.K.}
\and
\author{H.J. Walker}
\affil{CLRC, Rutherford Appleton Laboratory, Chilton, Didcot, Oxon OX11 0QX,
U.K.}

\begin{abstract}

Dust emission around the nearby star $\epsilon$ Eridani has been imaged
using a new submillimetre camera (SCUBA at the JCMT). At 850 $\mu{m}$
wavelength a ring of dust is seen, peaking at 60 AU from the star and
with much lower emission inside 30 AU. The mass of the ring is at least
$\sim$ 0.01 M$_{\earth}$ in dust, while an upper limit of 0.4
M$_{\earth}$ in molecular gas is imposed by CO observations. The total
mass is comparable to the estimated amount of material, 0.04--0.3
M$_{\earth}$, in comets orbiting the Solar System. 

The most probable origin of the the ring structure is that it is a young
analogue to the Kuiper Belt in our Solar System, and that the central
region has been partially cleared by the formation of grains into
planetesimals. Dust clearing around $\epsilon$ Eri is seen within the
radius of Neptune's orbit, and the peak emission at 35--75 AU lies within
the estimated Kuiper Belt zone of 30--100 AU radius. $\epsilon$ Eri is a
main-sequence star of type K2V (0.8 M$_{\odot}$)  with an estimated age of
0.5--1.0 Gyr, so this interpretation is consistent with the early history
of the Solar System where heavy bombardment occurred up to $\approx$~0.6
Gyr. An unexpected discovery is substructure within the ring, and these
asymmetries could be due to perturbations by planets.

\end{abstract}


\lefthead{Greaves et al.}
\righthead{Dust ring around $\epsilon$ Eri}

\section{Introduction}

One of the fundamental questions of astronomy is to determine how typical
the Solar System is. If Earth-like planets occur frequently, then life may
exist elsewhere in our Galaxy. However, the search for extra-terrestrial
planetary systems is extremely difficult, and is most often approached
indirectly. Massive planets introduce changes in the observed velocities
of their stars (the Doppler technique), and so far ten objects with
minimum masses up to 10 M(Jupiter) have been detected in this way (Mayor
\& Queloz 1995; Marcy et al.  1997).  Earth-like planets may be observable
during stellar transits; by ground and space-borne coronographs; or have
atmospheric signatures different from the stellar photospheres, effects
which could be detectable in the future (Fischer \& Pfau 1997). 

An alternative is to search for young systems where planets are still
forming from circumstellar material.  Molecular gas has been detected
around a few main-sequence stars, most of which are young objects
(Zuckerman, Forveille \& Kastner 1995; Dent et al. 1995). Dust was
detected around somewhat older main-sequence stars by IRAS (Aumann et al. 
1984), including the nearby systems Vega and Fomalhaut. These stars have
ages $\sim 10^8$ years, which lie at the end of the era when rocky planets
are expected to form.  Recently, submillimetre wavelength images of these
stars (Holland et al.  1998a) showed that Fomalhaut has a dust ring with a
central cavity, thus accumulation into planetesimals may have occurred. 
Such cavities have also been seen around $\beta$ Pic and HR4796A, at ages
of only $\sim 10^7$ years (Smith \& Terrile 1984; Jayawardhana et al. 
1998;  Koerner et al. 1998). Unexpected features are now emerging,
including secondary dust spots around Vega and $\beta$ Pic (Holland et al.
1998a), and these have yet to be explained by planet formation theories.

So far, there is little evidence for analogues of the Solar System around
stars of spectral type similar to the Sun. The Doppler technique is biased
towards massive planets close to their stars (current detections range out
to 2.5 AU), so it is difficult to find gas giants at the distances of
Jupiter and beyond. Submillimetre imaging can reveal much lower orbiting
masses of dust, below 1 M$_{\earth}$, but this emission is brightest when
the dust is heated by luminous stars. Vega, Fomalhaut and $\beta$ Pic are
all luminous A-type stars, with much shorter lifetimes than the Sun, so
that any planets would be short-lived. However, some G and K type stars
detected by IRAS are suited to ground-based follow-up; in particular 5
objects within 25 pc have significant 60 and 100 $\mu{m}$ excesses (Aumann
1988). Here we present the first image of dust around a low-mass
main-sequence star --- $\epsilon$ Eridani (HR 1084), a K2V
(0.8~M$_{\odot}$) star located only 3.22 pc from the Sun.  IRAS partially
resolved a warm dusty region (Gillett 1986), which is now shown to be a
ring similar in scale to the Kuiper Belt. The age of $\epsilon$ Eri is
estimated at $\leq$ 1 Gyr (Soderblom \& D\"{a}ppen 1989), so it may
represent a young analogue to the Solar System. 

\section{Observations and Results}

The observations were made with the new Submillimetre Common-User
Bolometer Array (SCUBA) (Holland et al. 1998b) at the James Clerk Maxwell
Telescope on Mauna Kea, Hawaii. The data were obtained between August 1997
and February 1998, using the SCUBA `jiggle-map' observing mode (Jenness,
Lightfoot \& Holland 1998). Fully-sampled maps were generated at 3$''$
sampling, with an on-source integration time of 12.1 hours. Although SCUBA
operates at 450 and 850 $\mu{m}$ simultaneously, this paper concentrates
mainly on the 850 $\mu{m}$ data, since observing conditions were generally
poor at 450 $\mu{m}$. Zenith atmospheric opacities at 850 $\mu{m}$ ranged
from 0.12 to 0.35 (only 30 \% of these data had 450 $\mu{m}$ opacities of
$<$~1) and calibration data were obtained from Mars and Uranus. Pointing
accuracy was 2$''$, small compared to the beam size of 15$''$ at 850
$\mu{m}$ (full-width at half maximum). The data were reduced using the
SCUBA User Reduction Facility (Jenness \& Lightfoot 1998), and are
rebinned in an RA-Dec. frame with 2$''$ cells.

\begin{figure}
\psfig{file=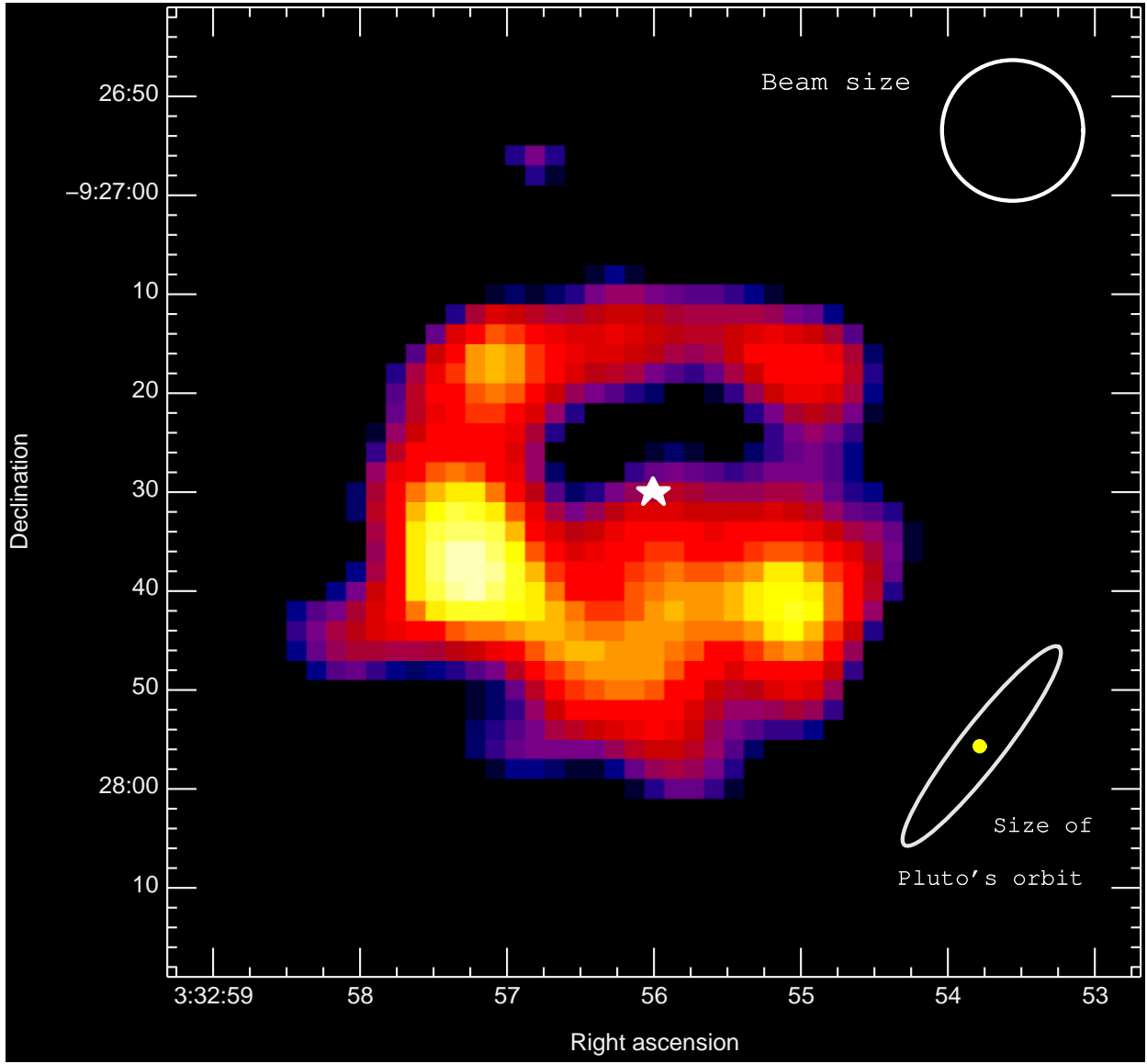,angle=-90,width=\textwidth}
\figcaption{ 
Dust emission around $\epsilon$ Eridani at a wavelength of 850 $\mu{m}$. 
The false-colour scale is linear from 2.8 mJy/beam (3.5$\sigma$ 
per pixel) to 8.5 mJy/beam (at the peak). The star is marked by the
asterisk symbol, the circle shows the 15$''$ beam size, and 1$''$
corresponds to 3.22 AU. The apparant size of Pluto's orbit at 3.22 pc 
distance is also shown. The position of the star is R.A. = 03h 32m 56.0s,
Dec. = $-09^{\circ} 27' 29.8''$, and is equinox 2000, epoch 1998. The
proper motion of the star was only 0.5$''$ over the 6 month observing
period. }
\label{fig1}
\end{figure}

The 850 $\mu{m}$ map of $\epsilon$ Eri is shown in Figure \ref{fig1}. The data have
been smoothed with a 8$''$ point-spread function, resulting in a peak
signal-to-noise per beam of 10. Photospheric emission of 1.7 $\pm$ 0.2 mJy
has also been subtracted in the image. The photospheric flux was estimated
by independent extrapolations using 2.2 and 3.4 $\mu{m}$ data (Carter
1990) and the IRAS 12 $\mu{m}$ flux (corrected for an effective
temperature of 5000 K). Dust emission around $\epsilon$ Eri was also
tentatively detected at 450 $\mu{m}$ (Table \ref{thetable}). 

\begin{deluxetable}{lccll}
\tablenum{1}
\tablewidth{0pt}
\tablecaption{Flux measurements for $\epsilon$ Eri}
\tablehead{
\colhead{Wavelength ($\mu{m}$)} & \colhead{Dust flux} & 
\colhead{Photospheric flux} & \colhead{Unit} & \colhead{Comments}\nl
}
\startdata
850 & 40 $\pm$ 3    & 1.7 $\pm$ 0.2 & mJy & r $\leq$ 35$''$ from star \nl
450 & 185 $\pm$ 103 & 6 $\pm$ 1     & mJy & r $\leq$ 35$''$ from star \nl
100 & 1.78 & 0.11 & Jy & IRAS \nl
60  & 1.34 & 0.29 & Jy & IRAS \nl
25  & 0.27 & 1.63 & Jy & IRAS \nl
12  & --- & 6.66  & Jy & IRAS \nl
3.4 & --- & 70.3  & Jy & SAAO \nl
2.2 & --- & 139.7 & Jy & SAAO \nl
1300 & ($>$ 7--24) & 0.7 & mJy & photometry, 11--24$''$ beams \nl
\enddata
\tablecomments{Flux data: this work, IRAS point-source catalogue, Carter
(1990), Zuckerman \& Becklin (1993), Chini et al. (1990, 1991).
Photospheric emission at 2.2 to 12 $\mu{m}$ was extrapolated to find dust
excesses (see also Gillett 1986). Only the 12 $\mu{m}$ point was used
for the IRAS wavelengths, and colour corrections were made for a 5000 K
photosphere (12--25 $\mu{m}$) and the dust spectral energy distribution
(60--100 $\mu{m}$).} 
\label{thetable}
\end{deluxetable}

The image shows extended flux around the star out to about 35$''$ radius.
The rest of the field of view is largely featureless, but one other source
is seen approximately 60$''$ east of $\epsilon$ Eri.  This has a total
flux of about 8 mJy, and could be a distant galaxy (e.g. Smail, Ivison \&
Blain 1997). Additionally, a 2.2 $\mu{m}$ image obtained with the Keck 10m
telescope shows three more background objects within 50$''$ of the star,
but none of these coincide with the submillimetre peaks. We conclude that
background sources are not significantly affecting the brightness
distribution seen around $\epsilon$ Eri at 850 $\mu{m}$. 

IRAS data at 60 $\mu{m}$ have shown a more compact dusty structure, with a
half-power radius of 8--11$''$ (Aumann 1991; Gillett 1986), and
temperatures ranging from 50 K up to 370 K for some grains near the star. 
The 850 $\mu{m}$ emission extends to about 35$''$ from $\epsilon$ Eri, and
only $\approx$~9~\% of it comes from the IRAS source area. We also find an
850/450 $\mu{m}$ flux ratio of 4.6 $\pm$ 2.6 (Table \ref{thetable}), corresponding to a
dust opacity index $\beta \leq 1.1$.  This implies large grains (Pollack
et al. 1994) which are efficient emitters and at the observed distances
from $\epsilon$ Eri will be at about 30 K (Backman \& Paresce 1993). These
results are discussed further below, and the fluxes from the millimetre to
the infrared are listed in Table \ref{thetable}.

A search was also made for a molecular gas component around $\epsilon$
Eri. Low limits had previously been set by a search for CO J=1--0
emission (Walker \& Wolstencroft 1988), using a large telescope beam
that included all the Figure \ref{fig1} emission region. More recently, Dent et
al. (1995) found no CO J=3--2 emission towards $\epsilon$ Eri, but the
14$''$ JCMT beam was only sensitive to the low-flux region near the star
(Figure \ref{fig1}). We therefore made a complementary search for CO J=2--1
emission, with a 21$''$ beam centred 3$''$ south of the bright peak in
Figure \ref{fig1}.  An upper limit of 25 mK was measured in a 1 km s$^{-1}$
spectral bin, which for a thermalised gas excitation temperature of
about 30 K corresponds to a CO column density of $\leq 1.4 \times
10^{13}$ cm$^{-2}$. CO is subject to photo-dissociation by interstellar
UV radiation, but accounting for this with standard models (van Dishoeck
\& Black 1988), the approximate column density of H$_2$ molecules is
$\leq 2 \times 10^{20}$ cm$^{-2}$ (A$_V \leq$ 0.2).

The total mass of gas and dust around $\epsilon$ Eri is estimated at less
than an Earth mass. For the dust component alone, we find 0.005--0.02
M$_{\earth}$ (1.5--6 $\times 10^{-8}$ M$_{\sun}$), assuming T$_{\rm dust}$
of 30 K and an absorption coefficient $\kappa_{\nu}$ between 1.7 and 0.4
cm$^2$ g$^{-1}$. The lower value of $\kappa_{\nu}$ is suggested by models
of large, icy grains (Pollack et al. 1994), while the higher estimate has
been used for previous observations of Vega-type stars (e.g. Holland et
al. 1998a). However, very large grains could dominate the mass while
adding little emission, so both mass estimates are lower limits. The CO
upper limit corresponds to $\leq$ 0.4 M$_{\earth}$ of molecular gas, when
extrapolated over the area of dust emission above half-maximum brightness. 

\section{Discussion}

The 850 $\mu{m}$ image shows extended emission, which most resembles a ring
of dust around the star. The emission peaks at a radius of 18$''$ ($\approx$
60 AU), and substantially reduced emission is seen at about 30 AU. This
ring-like morphology was previously suspected from single-beam observations
at 1.3 mm, where flux variations were found with different beam sizes
(Zuckerman \& Becklin 1993), but is only now seen directly. The image also
shows some surprising asymmetries and bright peaks, which are discussed
further below. 

\begin{figure}
\psfig{file=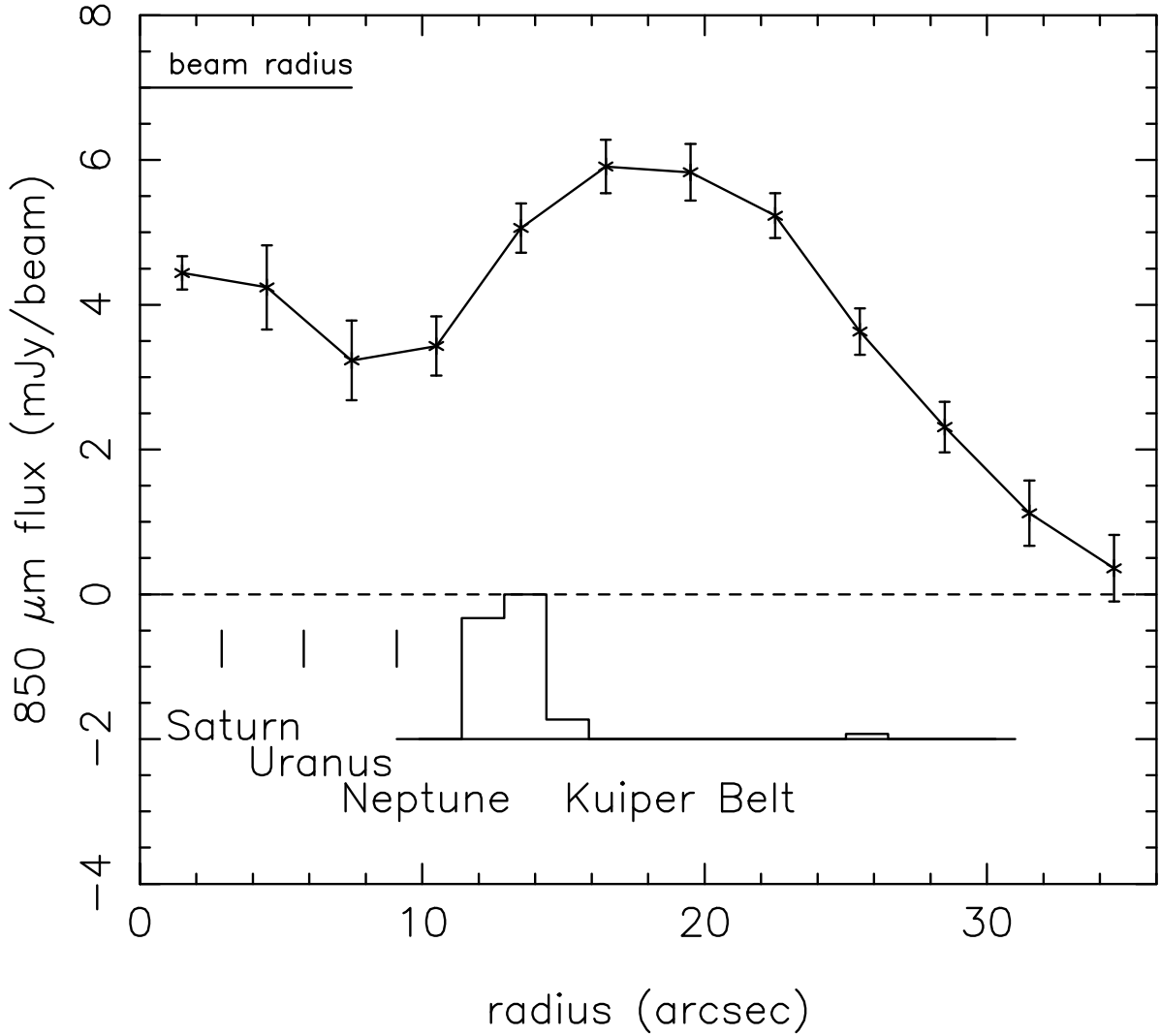,angle=-90,width=\textwidth}
\figcaption{
Radial profile of dust emission around $\epsilon$ Eri.
The mean 850 $\mu{m}$ flux density (in mJy/beam) is plotted against radial
distance from the star. The data are averaged in 3$''$ bins, to match the
raw image sampling, and the error bars represent the standard error of the
mean from the dispersion of signals at that radius.  For comparison, the
number of known Kuiper Belt objects is plotted underneath as a function of
semi-major axis (Jewitt 1997), together with the locations of the outer
planets.
}
\label{fig2}
\end{figure}

Figure \ref{fig2} shows the azimuthally averaged radial profile calculated from the
850 $\mu{m}$ map. There is a difference of a factor of two between the
flux densities of the ring (r = 18$''$) and the cavity minimum (r
$\approx$ 8$''$). The cavity region out to this distance has only 6 \% of
the total integrated flux. There is an apparent rise in flux density
within r = 4.5$''$, which in the unsmoothed data appears to correspond to
a small peak south of the star. However, the signal level in this region
is uncertain, as it includes less than 10 map pixels.

The most probable explanation for the ring-like structure is a young
analogue of the Kuiper Belt. The central deficit of emission suggests
accumulation of dust into planetesimals (which emit much less per unit
mass than individual grains). The $\epsilon$ Eri system could thus be
analogous to the young Solar System, seen when planet formation is ongoing
or complete, but some dust is still present at all radii out to about
36$''$ (115 AU). The age of the star is not well-defined, but can be
estimated from the level of chromospheric activity. An age relation for
dwarf stars is discussed by Soderblom, Duncan \& Johnson (1991) and their
results for $\epsilon$ Eri (Soderblom \& D\"{a}ppen 1989) suggest an age
of $\approx$ 0.5--1.0 Gyr. At this age, a partially-cleared but still
dusty system is seen as expected.

Figure \ref{fig2} also shows the locations of the Solar outer planets and the
Kuiper Belt, sketched below the dust profile of $\epsilon$ Eri. The
least dust emission is seen at the equivalent of Uranus' and Neptune's
orbits, while the ring peaks at about 60 AU, well within the Kuiper Belt
zone. Since the 850 $\mu{m}$ image is broadened by the beam size of
15$''$, the dust ring must in fact be narrower than it appears in Figure
\ref{fig2}. The observed width at half-maximum is from radii of 7.5$''$ to
27$''$, which when deconvolved from the beam imply half-maximum points
at about 11$''$ and 23$''$, or 35--75 AU. This is similar to our Solar
System, as the Kuiper Belt has an inferred declining density outwards
from r $\approx$ 32 AU (Jewitt \& Luu 1995), with much of the population
within 100 AU.

IRAS observations suggested considerable emission from the cavity region,
with warm grains (50--370 K) inside 8--11$''$ radius (Aumann 1991; Gillett
1986). At 850 $\mu{m}$, this inner region contributes only $\sim$~9~\% of
the flux, and for the higher T$_{\rm dust}$, $<$ 5~\% of the overall mass.
Thus there is a genuine cavity within the ring, although it does contain
some grains.  In fact, the inner region is much dustier than the Sun's
zodaical belt, by two to three orders of magnitude (Zuckerman \& Becklin
1993). This persistence of central material around $\epsilon$ Eri at
0.5--1.0 Gyr is consistent with the history of the Solar System, where the
Earth was heavily bombarded by asteroids up to about 0.6 Gyr (Maher \&
Stevenson 1988) . 

Other mechanisms can reduce the amount of dust emission near a star,
including grain mantle sublimation or radiation-grain drag
(Poynting-Robertson effect). However, ice mantles should persist to within
a few AU of $\epsilon$~Eri, since the giant planets in the Solar System
are believed to have formed around rock and ice cores, and the luminosity
of $\epsilon$ Eri is only about 0.33 L$_{\sun}$ (Soderblom \& D\"{a}ppen
1989). For P-R drag, grains $\sim$ 1 mm in diameter would have been
cleared only to radii of about 15 AU (Jura 1990), even if the star is as
old as 1 Gyr. Thus it would be difficult to reproduce clearing out to the
observed 35 AU. Also, P-R drag naturally produces a 1/r density
distribution as small grains spiral in towards the star, and this is not
seen. Thus it appears unlikely that sublimation of grain mantles or
radiation-grain drag can explain the reduced emission close to $\epsilon$
Eri. 

The total mass of the $\epsilon$ Eri ring depends on the presence of gas,
but is at least $\sim$ 0.01 M$_{\earth}$, and possibly as high as 0.4
M$_{\earth}$. This range encompasses the present-day mass of comets in the
Kuiper Belt, $\approx$ 0.04--0.3 M$_{\earth}$ between 30 and 100 AU
(Backman, Dasgupta \& Stencel 1995). The $\epsilon$ Eri circumstellar ring
could therefore be evolving into a Kuiper Belt analogue of icy cometary
bodies. Comets may form in only a few 10$^5$ years (Weidenschilling 1997),
but the time to accrete and/or disperse all the dust is likely to be $\gg
10^8$ years, the period estimated to form large Kuiper Belt objects
(Kenyon \& Luu 1998). 

The $\epsilon$ Eri ring system appears close to face-on, as the morphology
is roughly circular (Figure \ref{fig1}). The major:minor axis ratio is $\sim$ 1.1 (66
and 60$''$ diameters at the 3$\sigma$ level), implying an inclination to the
plane of the sky of i $\sim$ 25$^{\circ}$. This is in good agreement with
the orientation deduced for the stellar pole from optical line data, of
i~$\approx$~30 $\pm$ 15$^{\circ}$ (Saar \& Osten 1997). These results
suggest that the dust ring is aligned with the stellar equator.

A surprising result is the non-uniformity of the ring
(Figure \ref{fig1}). The bright peak seen at dRA,dDec. = (+19$''$,--7$''$) has a
flux density of 6.9 mJy in a 15$''$ beam area, compared to the region on
the opposite side of the star where the flux density is only 3.5 mJy per
beam, a difference of five times the rms noise. The bright peak is a real
feature (identified in 13 out of 16 of the individual maps). The fainter
peaks north-east and south-west of the star are less certain, as they were
identified in only half of the individual maps. 

An enhancement of dust density might represent the wake of a planet orbiting
within the ring. Dust can become trapped in resonant orbits with a planet,
and for example, Dermott et al. (1994) found that up to 20 \% of the
zodiacal dust near the Earth is trapped to form a dust condensation, seen in
IRAS data. Alternatively, a planet orbiting just inside the $\epsilon$ Eri
ring could cause transient features, similar to the manner in which Neptune
`erodes' the inner edge of the Kuiper Belt (Jewitt \& Luu 1995). Asymmetries
can also be produced when the peculiar velocity of the star with respect to
the interstellar medium causes interstellar grains to stream into the disk,
and erosion effects produce a brightness change on one side (Artymowicz \&
Clampin 1997). This process is likely to be effective only at very large
radii (hundreds of AU). Finally, Backman \& Paresce (1993) have pointed out
that planetesimal collision rates are low at a few tens of AU, thus recent
collisions will produce discontinuous features. The flux enhancement at
(+19$''$,--7$''$) is 2.6 mJy, or the amount of dust that would be produced
by the destruction of a body with 15--55~\% of the mass of Pluto (for
T$_{\rm dust}$ = 30 K, $\kappa_{\nu}$ = 1.7--0.4 cm$^{-2}$ g$^{-1}$). 
However, it is unlikely that we would observe such a major collision, as the
remnants would disperse within a few orbital periods (500 years at 60 AU),
or a fraction $<$~10 $^{-5}$ of the age of the system. 

\section{Conclusions}
 
A dust ring has been detected around $\epsilon$ Eri with a mass and radius
similar to the Kuiper Belt. Dust appears to be partially cleared inside
the ring, consistent with accumulation into planetesimals. Other
mechanisms of producing decreased dust emission cannot be definitively
ruled out, but Poynting-Robertson drag and grain mantle sublimation appear
unlikely to produce such a cavity. An inhomogeneity in the ring suggests
the presence of a large orbiting body, either within the ring or just
inside it. This corresponds to an orbit similar to Neptune's, or up to
twice as large. 

Any planet formation around $\epsilon$ Eri is probably complete, as the
stellar age is around 0.5--1.0 Gyr, and Earth-like planets are believed to
form within 0.1 Gyr (gas giants on shorter timescales). No direct evidence
has as yet been found for planets around $\epsilon$ Eri.  Indications of a
10-year variation in radial velocity (Walker et al. 1995) are now
suspected to be a multiple of a 5 year period in changes in the stellar
photosphere (Gray \& Baliunas 1995).  The amplitude of the suspected
radial velocity was $\sim$ 15 m s$^{-1}$, and more recent searches (G.
Marcy, priv.  comm.) find no such effect, with errors of 10--15 m s$^{-1}$
over a 10 year period. However, the radial velocities will be reduced
because the $\epsilon$ Eri system is seen almost face-on, making it harder
to detect Jovians by this method.  The new submillimetre image suggests
that the $\epsilon$ Eri system is a strong candidate for an analogue of
the young Solar System, and that renewed planetary searches by various
techniques could be rewarding.

\acknowledgements
{The JCMT is operated by the Joint Astronomy Centre, on behalf of the UK
Particle Physics and Astronomy Research Council, the Netherlands
Organisation for Pure Research, and the National Research Council of
Canada. This research was supported in part by PPARC funding and by NSF
and NASA grants to UCLA. }

\end{document}